\documentclass[preprint,twocolumn,3p]{elsarticle}
\usepackage[colorlinks,linkcolor=blue,anchorcolor=blue,citecolor=blue]{hyperref}
\usepackage{subfigure}
\usepackage{booktabs}
\usepackage{extarrows}
\usepackage{balance}


\newcommand{\grad}{\ensuremath{^{\circ}}}

\begin{document}
\bibliographystyle{unsrt}

\title{Fabrication of \(CeO_2\) by sol-gel process based on microfluidic technology as an analog preparation of ceramic nuclear fuel microspheres\tnoteref{t1}}
\tnotetext[t1]{Funded by National Nature Science Foundation of China (91226109 \& 21076203)}

\author[snst]{Bin Ye}
\author[snst]{Jilang Miao}
\author[snst]{Jiaolong Li}
\author[snst]{Zichen Zhao}
\author[snst]{Zhenqi Chang\corref{cor1}}
\ead{zqchang@ustc.edu.cn}
\author[G2IP]{Christophe A. Serra}

\cortext[cor1]{Corresponding author}
\address[snst]{School of Nuclear Science and Technology, University of Science and Technology of China, Huang-Shan Road, He-Fei, P. R. China}
\address[G2IP]{RGroupe d'Intensification et d'Intrapolation des Proc\'{e}d\'{e}s Polym\`{e}res (G2IP), Laboratoire d'Ing\'{e}nierie des Polym\`{e}res pour les Hautes Technologies (LIPHT) – EAc(CNRS) 4379, \'{E}cole Europ\'{e}enne de Chimie, Polym\`{e}res et Mat\'{e}riaux (ECPM), Universit\'{e} de Strasbourg (UdS), 25 rue Becquerel, F-67087 Strasbourg, France}


\begin{abstract}
Microfluidics integrated with sol-gel processes is introduced in preparing monodispersed MOX nuclear fuel microspheres using nonactive cerium as a surrogate for uranium or plutonium. The detailed information about microfluidic devices and sol-gel processes are provided. The effects of viscosity and flow rate of continuous and dispersed phase on size and size distribution of \(CeO_2\) microspheres have been investigated. A comprehensive characterization of the \(CeO_2\) microspheres has been conducted, including XRD pattern, SEM, density, size and size distribution. The size of prepared monodisperse particles can be controlled precisely in range of \(10\mu m\) to \(1000\mu m\) and the particle CV is below 3\%.
\end{abstract}
\maketitle

\balance
\section{Introduction}
As an unavoidable by-product of the current uranium fuel cycle in thermal reactors, there is an excess of plutonium inventory around the world \cite{1}. Besides plutonium, the current fuel cycle in thermal reactors or fast reactors around the world annually produces great amount of long-lived minor actinides (MA), including neptunium, americium and curium \cite{1}. For the sakes of their radio toxicity, generation of decaying heat, very long half-lives and applicability of nuclear weapons, MA elements are a major concern for environmental safety \cite{1}. 

There are many strategies for the disposal of Pu and MA elements, among which the approach of transmutation is considered the effective and economic solution \cite{1}. The sub-critical accelerator driven system (ADS) is being considered as a potential means of transmutation \cite{2}.Compared with other types of nuclear fuel, dispersed nuclear fuel is fit for ADS. There are two kinds of promising nuclear fuels for ADS, CERMET and CERCER composites \cite{3}. The composite targets contain \(PuO_2\) pellets, MA elements and inert matrix materials. The EUROPEANS Integrated Project has stated that the CERMET with the Mo-matrix is recommended as the reference fuel and CERCER with the MgO matrix as a back-up solution \cite{2}.

The fabrication of CERMET composite pellets is based on the fabrication of particles containing the actinide phase by a combination of the internal gelation \cite{4,5}, gel supported precipitation and the infiltration methods, followed by mixing the particles and the matrix powder by conventional blending methods. The main advantage of such composite targets compared to a solid solution is that it potentially minimizes irradiation-induced property changes in the fuel pellets by localizing the fission damage in a limited geometric domain within the fuel \cite{6}.The sol-gel processes which can directly convert plutonium and MA elements from solution to consolidated formulations possesses the most brilliant future, since handling compounds of MA elements in powder form is very difficult because of their radio toxicity \cite{7}. Particularly, the sol-gel process steps are rather simple and no excessive liquid waste is produced, which is a major concern in the selection of the fabrication process. The highly porous beads of \(PuO_2\) were produced by the external gelation method as porous medium. After their calcinations, the beads were infiltrated with MA nitrate solutions to reach required MA content followed by their mixing with the Mo-matrix \cite{2}.

But the fabrication of composite targets is considerably more difficult than that of solid solution oxide pellets. This is a result of the specific requirements for size distribution of the dispersed pellets. The volume of ceramic phase and its distribution thus play a significant role in the case of fabrication \cite{6}. 

The current sol-gel process has limitation in maintaining suitable homogeneity of size-distribution. Recently capillary-based microfluidic devices were adopted for the production of microspheres \cite{8}. Microfluidic device offers precise control over the flows of the fluids at small scales \cite{9}. This method apparently exploits a shear-rupturing mechanism in which the size of the droplets is controlled by the capillary number (\(Ca=ν\nu/\gamma\), where v is the velocity of the continuous liquid, \(\nu\) its viscosity, and \(\gamma\) the interfacial tension)—that is, by the interplay of the shear stress and surface tension. It has practical applicability value in the fabrication of fuel pellets \cite{9}.

In this study, we focus on the preparation of microspheres for CERMET composite fuel. The microspheres were supposed to be \(PuO_2\), but due to the radioactivity of Pu, we have used Ce as a surrogate for Pu \cite{10}. Related researches have proved that \(CeO_2\) has similar sintering behavior, pore former effect and thermal properties with \(PuO_2\) \cite{10}. A flexible microfluidic device was designed and used for synthesizing mono-disperse highly porous beads of \(CeO_2\).The diameter of the microspheres can be precisely controlled at the range of \(10\mu m\) to \(1000\mu m\) by adjusting experimental parameters. To the nuclear element microspheres-dispersed nuclear fuel, the monodispersity and precise size controllability of the nuclear fuel microspheres is beneficial to form uniform nuclear element distribution in the composite fuels or targets. 

\section{Experiments}
\subsection{Microfluidic device}
\begin{figure}
\centering
\includegraphics[width=0.45\textwidth]{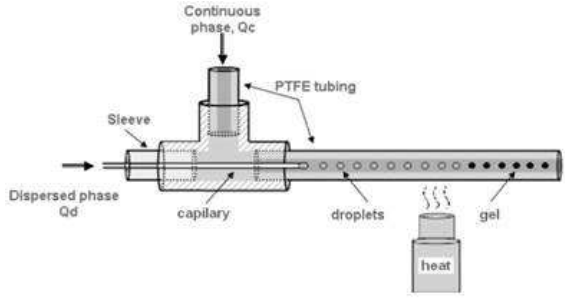}
\caption{Radial line power density for the core}
\label{fig_sch}
\end{figure}

A schematic drawing of the microfluidic system is shown in Fig\ref{fig_sch} \cite{14}.A capillary was inserted inside a T-junction (P-728-01, Upchurch Scientific) along its main axis. Two syringe pumps (LSP01-1A, Longer Pump) were used to deliver the continuous and monomer phases at a specific flow rate. The dispersed phase flows in through the left junction, the continuous phase through the top junction. And when the two phases flow out through the right junction, droplets were sheered and thus gels were formed.

Experiments reported here were run using capillaries with outer/inner diameter of \(385\mu m\)/\(251\mu m\) and PTFE tubing with outer/inner diameter of 1.6mm/1.06mm.

Direct observation of the droplet formation was made by coupling a CCD camera (uEye UI-2220SE, IDS) with a microscope (XSP-30E, Shanghai halibut instrument limited company). The camera captures up to 52 fps at a full resolution of 768\(\times\)576 pixels.

\subsection{Materials}
Dimethyl silicone oil (AR, Sinopharm Chemical Reagent Co., Ltd) was used as continuous phase in the microfluidic device. And an equimolar mixture of urea \((CO(NH_2 )_2)\) (AR, Shanghai Suyi Chemical Reagent Co., Ltd) and HMTA \(((CH_2 )_6 N_4)\) (AR, Tianjin Guangfu Science and Technology Development Co., Ltd) dissolved in distilled water, then mixed with \((NH_4)_2Ce(NO_3)_6\)  (AR, Sinopharm Chemical Reagent Co., Ltd) solution in the certain mole ratio(R) 2.3\cite{12} was used as the dispersed phase.  The inner phase ingredient is presented in Table \ref{tabDPhI}. 

Carbon tetrachloride \(CCl_4\)  (AR, Sinopharm Chemical Reagent Co., Ltd) and ammonia (AR, Sinopharm Chemical Reagent Co., Ltd) solutions were used to wash the gel particles.

\begin{table}
\caption{Dispersed phase ingredient}
\begin{center}
\begin{tabular}{lll}
\toprule
Chemical & Amount(mol) & volume(ml)\\
\midrule
Urea  & 0.014  &     \\
HMTA  & 0.014  &     \\
\((NH_4)_2Ce(NO_3)_6\) & 0.006    &       \\
Distilled water &       &  10 \\
\bottomrule
\end{tabular}
\end{center}
\label{tabDPhI}
\end{table}

As can be seen from Table \ref{tabDPhI}, the mole ratio(R) of HMTA-urea/Ce is 2.3 and the solution was close to saturated solution. The saturated solution indicates the least water in the gels and ensures their stability and good sphericity.

\subsection{Fabrication Process}
In our experiment, ammonium ceric nitrate is mixed with urea and HMTA solution under cooled conditions (5\grad). The transparent mixture, the inner phase is injected through the capillary from the left inlet of the T-junction as shown in Fig\ref{fig_sch} The silicone oil, the continuous phase flows in through the top inlet.

\begin{figure}
\centering
\subfigure[droplet formation]{\includegraphics[width=0.21\textwidth]{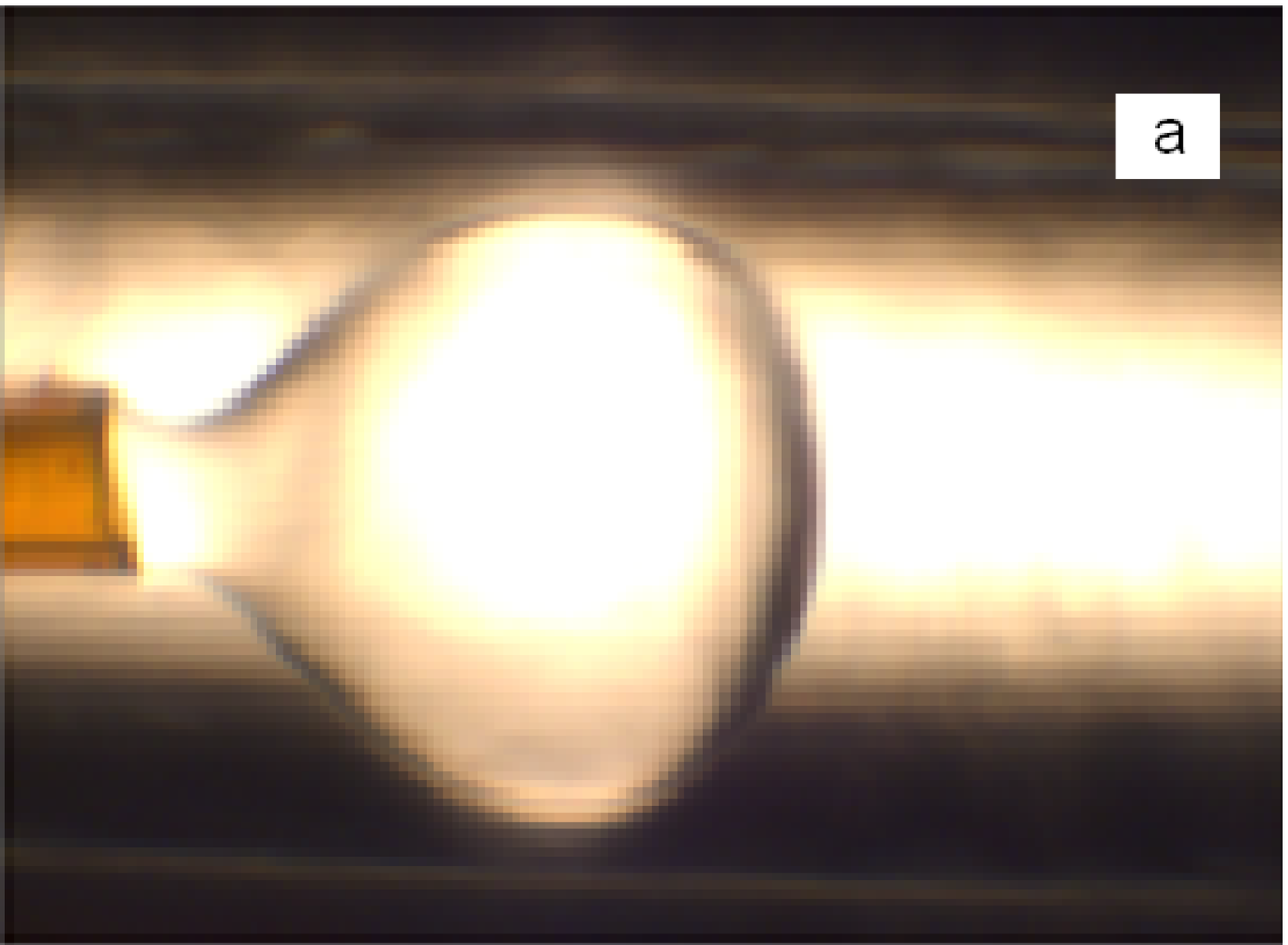}}\label{ffma}
\subfigure[gels, scale bar \(200\mu m\)]{\includegraphics[width=0.21\textwidth]{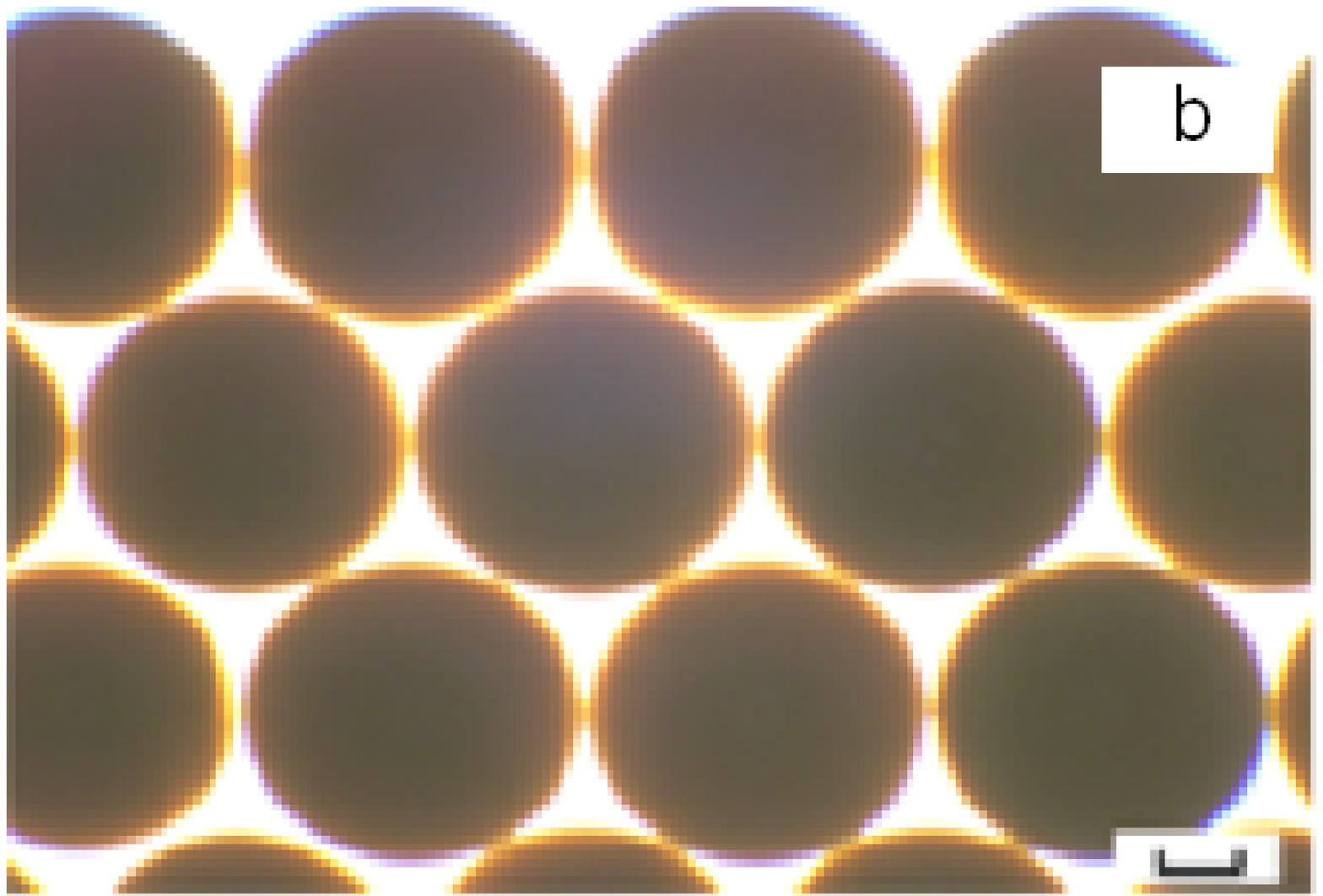}}\label{ffmb}
\caption{Optical microscope image of microspheres formation}
\label{fig_fm}
\end{figure}

As shown in Fig\ref{ffma}, the inner phase droplet is formed at the outlet of the capillary by the sheering of the continuous phase. The formed droplets are contacted with the silicone oil and become gel microspheres after being heated (~90 \grad), which is given in Fig\ref{ffmb} \cite{4}

\begin{figure*}[!t]
\begin{eqnarray}
(CH_2)_6N_4 + H^+ &=& ((CH_2)_6N_4)H^+ \label{eq1}\\
((CH_2)_6N_4)H^+ +9H_2O &=& 6HCHO + NH_4^+ +3NH_4OH \label{eq2}
\label{eq}
\end{eqnarray}
\end{figure*}

\begin{figure}
\centering
\includegraphics[width=0.39\textwidth]{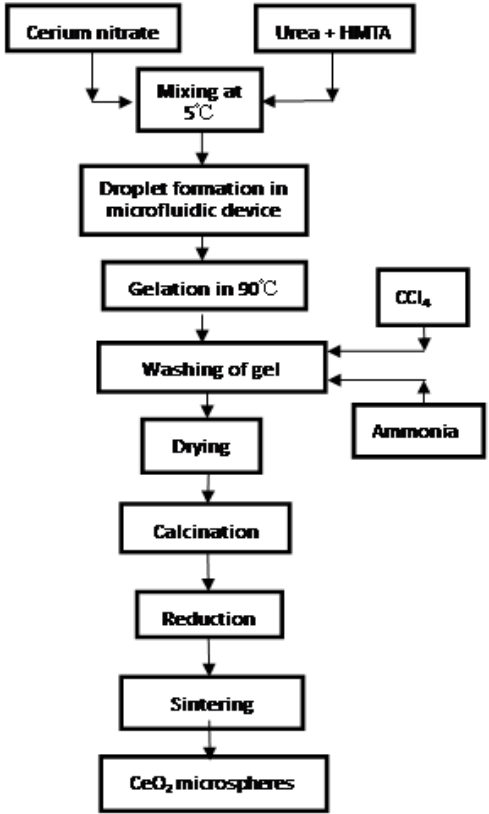}
\caption{Radial line power density for the core}
\label{fig_fc}
\end{figure}

These gel microspheres are washed first with \(CCl_4\) to remove the silicone oil and then with \(NH_4OH\) solution to remove excess gelation agents HMTA, Urea and ammonium nitrate. The washed microspheres are dried at 120 \grad in air and then calcined up to 500\grad to remove residual organic matter and ammonium nitrate. The calcined microspheres are then reduced at 600\grad. The \(CeO_2\) microspheres thus produced are sintered at 900\grad for 2 hours of theoretical density (TD)\cite{4}. Flowchart for the above fabrication process is shown in Fig\ref{fig_fc}. 

\subsection{Characterization}
The crystallographic phase analysis was conducted with X-ray diffraction (XRD) (MXPAHF, 18KW) at room temperature using Cu \(K_{\alpha}\) radiation. The microstructure of the \(CeO_2\) pebbles was observed with scanning electron microscope (FE-SEM SUPRA 55). The open pore size distribution of the sintered microspheres was measured by a Surface Area and Porosimetry Analyzer (V-sorb 2800). 

\section{Results and discussion}
\subsection{fabrication of \(CeO_2\) in sol-gel process}
Sol–gel process provides an alternate route for fabrication of ceramic nuclear fuel. In comparison with the conventional powder pellet fabrication process, the sol–gel process possesses several advantages by eliminating handling of radioactive powders. The sol–gel process uses only fluids or fluid like materials, thus become amenable to remote handling \cite{4}.

V. N. Vaidya has described the principle of reaction among urea, HMTA and heavy metal ions in sol-gel process \cite{4}. In this experiment, the feed solution is made by mixing heavy metal nitrate solution with equimolar mixture of urea \(CO(NH_2)_2\) and HMTA \((CH_2)_6N_4\) under cooled conditions (5\grad). HMTA undergoes hydrolytic decomposition releasing ammonia (reaction \eqref{eq1}\eqref{eq2}), which causes the gelation of cerium hydrated oxide under heating processes

And urea plays an important role by preventing premature gelation and consuming formaldehyde to form methylol urea derivatives that enhance the hydrolytic reaction \cite{13}.

Gelation was achieved by heating droplets of this broth by contact with hot (90\grad) Dimethyl silicone oil. The reaction was driven forward by scavenging of H+ ion with HMTA (reaction \eqref{eq1})\cite{7}.
Whether metal ions of \(Ce^{4+}\) reacts as described by V.N. Vaidya has not yet been confirmed, but with X-ray diffraction it is certain that the \(CeO_2\) microspheres were finally obtained by sintering the gel microspheres.

\begin{figure}
\centering
\includegraphics[width=0.45\textwidth]{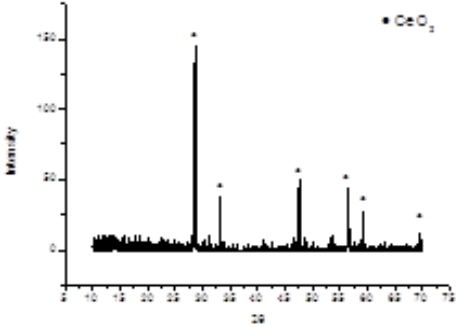}
\caption{X-ray diffraction pattern of the sintered microsphere}
\label{fig_xrd}
\end{figure}

The sintered particle XRD experiments were carried out using a diffractometer with Cu \(K_{\alpha}\) radiation (\(1.54056 \mathring{A}\)). And the XRD pattern is shown in Fig\ref{fig_xrd}. 

From XRD experiments data as shown in Fig\ref{fig_xrd}, the sintered particles have been identified as cubic phase \(CeO_2\) with lattice parameter a being \(5.40\mathring{A}\). Therefore, it is certain that the fabrication process given in Fig\ref{fig_fc} results in \(CeO_2\) successfully. 

\subsection{Controllable size of \(CeO_2\) microsphere in microfluidic device}\label{scont}
The process of preparing gelation particles in microsystem is that the sol droplets are first produced in the microfluidic devices and then solidified downstream by gelation. The uniform droplet formation is the basis of synthesizing uniform nuclear fuel particles.The droplet size can be controlled by regulating experiment parameters such as flow rate and viscosities of disperse and continuous phases.

\begin{figure}
\centering
\includegraphics[width=0.48\textwidth]{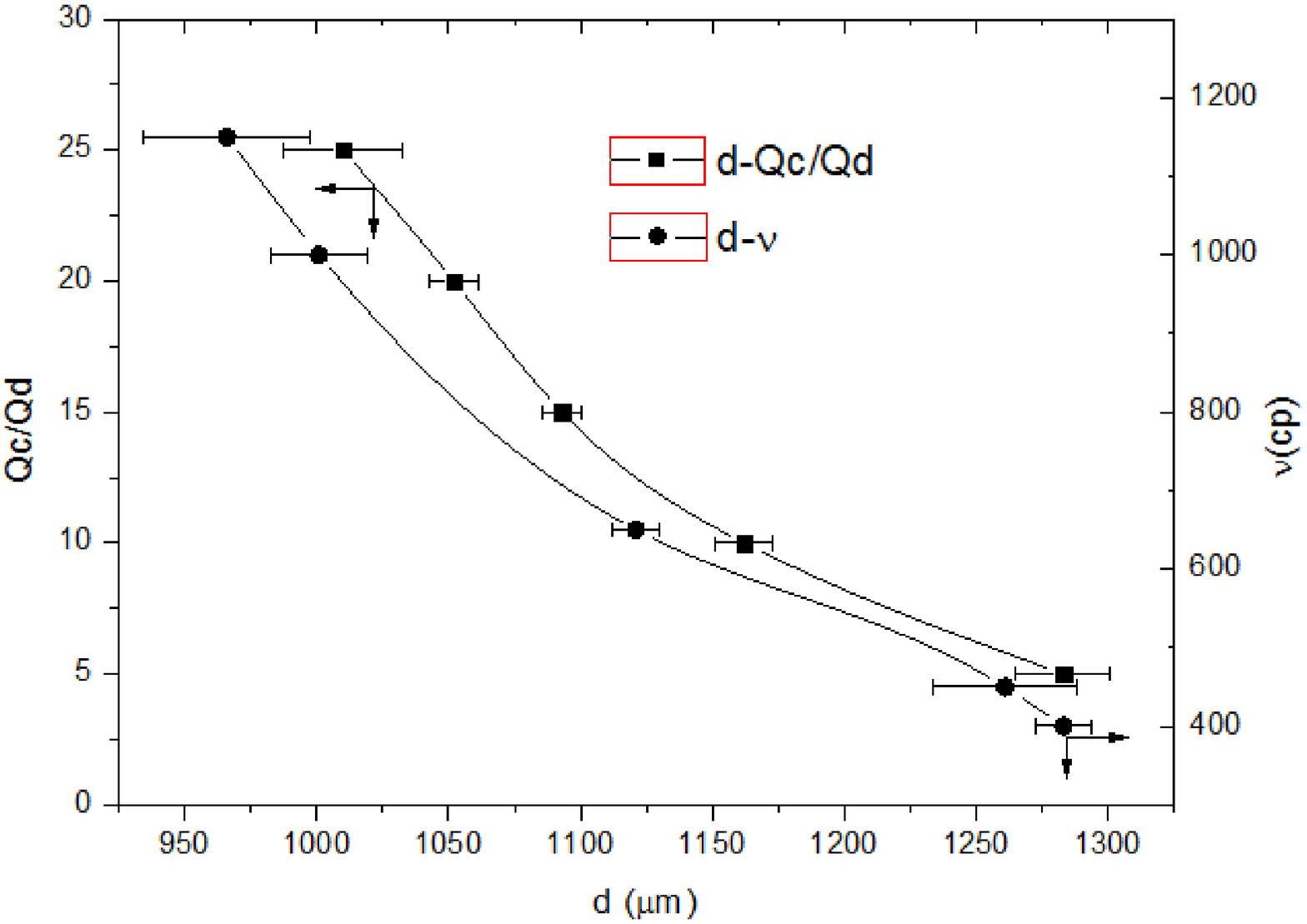}
\caption{The influence of flow rate ratio and viscosity on size of gels (\(Q_c\): flow rate of continuous phase; \(Q_d\): flow rate of dispersed phase; \(\nu\): viscosity of the continuous phase)}
\label{fig_rl}
\end{figure}

As seen from the \(d-Q_c/Q_d\) curve in Fig\ref{fig_xrd}, microsphere size decreases with the ratio of flow rate of continuous phase to that of the dispersed one. High continuous flow rate \(Q_c\) contributes to the size decrease in that the droplet detachment is advanced namely the breakup time is shorter since the high continuous phase velocity raises drag force and makes the equilibrium between the interfacial tension force and hydrodynamic force reached earlier. In addition, \(Q_d\) results in larger microsphere size in a more direct way. Under ordinary experiment conditions, the breakup time is mainly determined by the continuous flow rate but remains approximately constant for varied disperse phase velocity \cite{16}. Therefore, the increase in dispersed phase flow rate leads more dispersed phase fluid to accumulate in the capillary tip under the same droplet breakup time, which forms a larger droplet size.

Besides flow rates, the fluid viscosity also strongly affects the droplet formation. The relationship can be seen from the \(d-\nu\) curve in Fig\ref{fig_rl} that the higher the viscosity of the continuous phase fluid is, the smaller the droplet diameter is. The reasons are that the shorter droplet breakup time is introduced by increasing the sheer force exerted on the continuous phase.

These results show the predictability of the gel size namely the possibility of adjusting process parameters and material properties to obtain desired gel diameter.  In a previous similar work, Zhengqi Chang has discussed the relation and summarized the empirical formula \cite{8}.The size-controlled preparation of the feed droplet, namely the desired gel microsphere diameter can be obtained by regulating the experimental parameter.

\subsection{Characterization of microsphere features}
\subsubsection{The density of the sintered microspheres}
The density is calculated to be \(6.05g/cm^3\), the theoretical density of \(CeO_2\) is \(7.13 g/cm^3\), the difference between real and theoretical density can be attributed to the pores formed in the microspheres. 

\subsubsection{Expected narrow size distribution}
\begin{figure}
\centering
\includegraphics[width=0.36\textwidth]{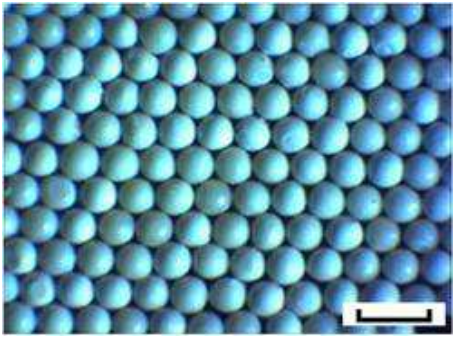}
\caption{Microscope image of the fabricated particles, scale bar \(400\mu m\)}
\label{fig_msarr}
\end{figure}

As described in section\ref{scont}, the size of the fabricated particles can be controlled with selected material properties and procedure parameters. The microspheres after being sintered are shown in Fig\ref{fig_msarr}. The mean diameter of a microsphere given in Fig\ref{fig_msarr} is \(200\nu m\). Particles with size about \(200\nu m\) suffice to realize the aim of localizing damages within small volume for fuel pins in the scale of 1 to 10 mm. Pellets containing small microspheres avoid large temperature gradients in large microspheres and thus distribute the heat to the inert matrix quickly.

In addition, a narrow microsphere size distribution is achieved with the microfluidic device. The narrow distribution is validated with the small CV value less than 3\%.  A narrow microsphere size distribution contributes to a better in-core behavior in two ways. First, a wide size distribution makes it difficult to blend the fuel particles and inert matrix powders sufficiently. The potential large pores and cracks in the pellets undermine the mechanical properties of fuel elements. Secondly, homogenous fuel particle distribution leads to a similarly well-distributed reaction rate, and consequently uniform heat deposition and radiation damage.

\subsubsection{Microstructure morphology and porosity of \(CeO_2\)}
\begin{figure}
\centering
\subfigure[sphericity characterization, scale bar \(200\mu m\)]{
\label{fig_SEM1}
\includegraphics[width=0.215\textwidth]{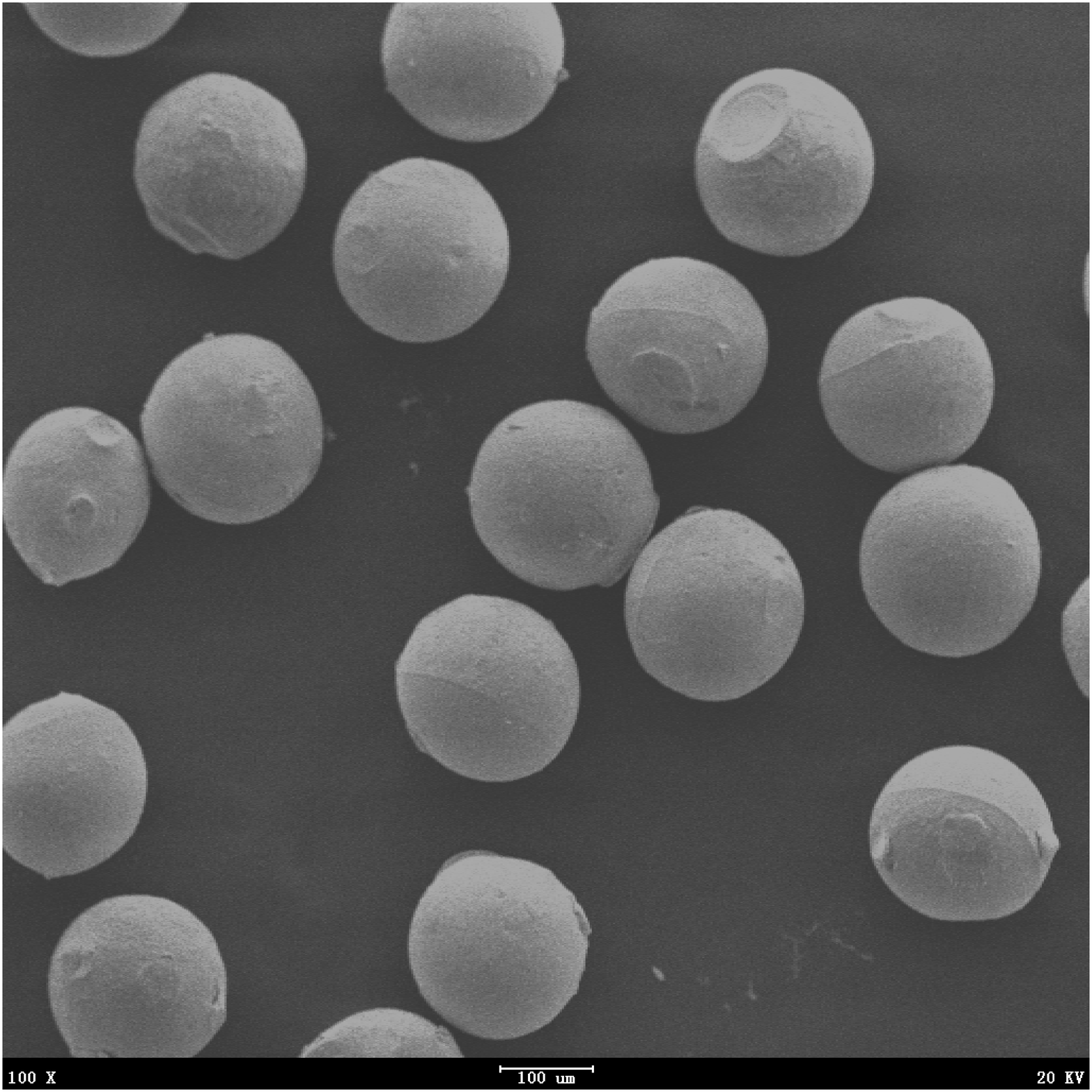}}
\subfigure[sphericity characterization, scale bar \(200\mu m\)]{
\label{fig_SEM2}
\includegraphics[width=0.215\textwidth]{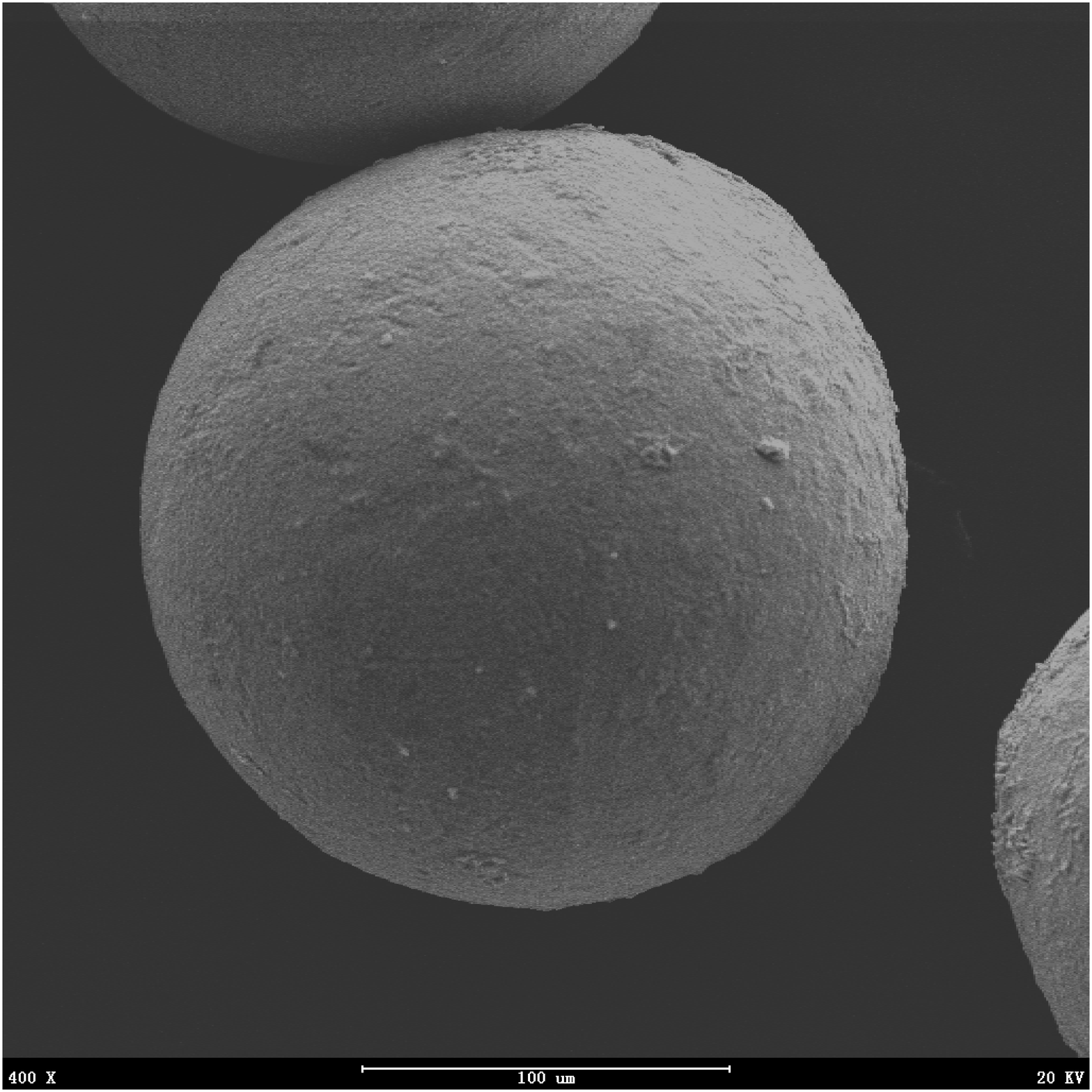}}
\\
\subfigure[ porous surface exhibition, scale bar \(10\mu m\)]{
\label{fig_SEM3}
\includegraphics[width=0.45\textwidth]{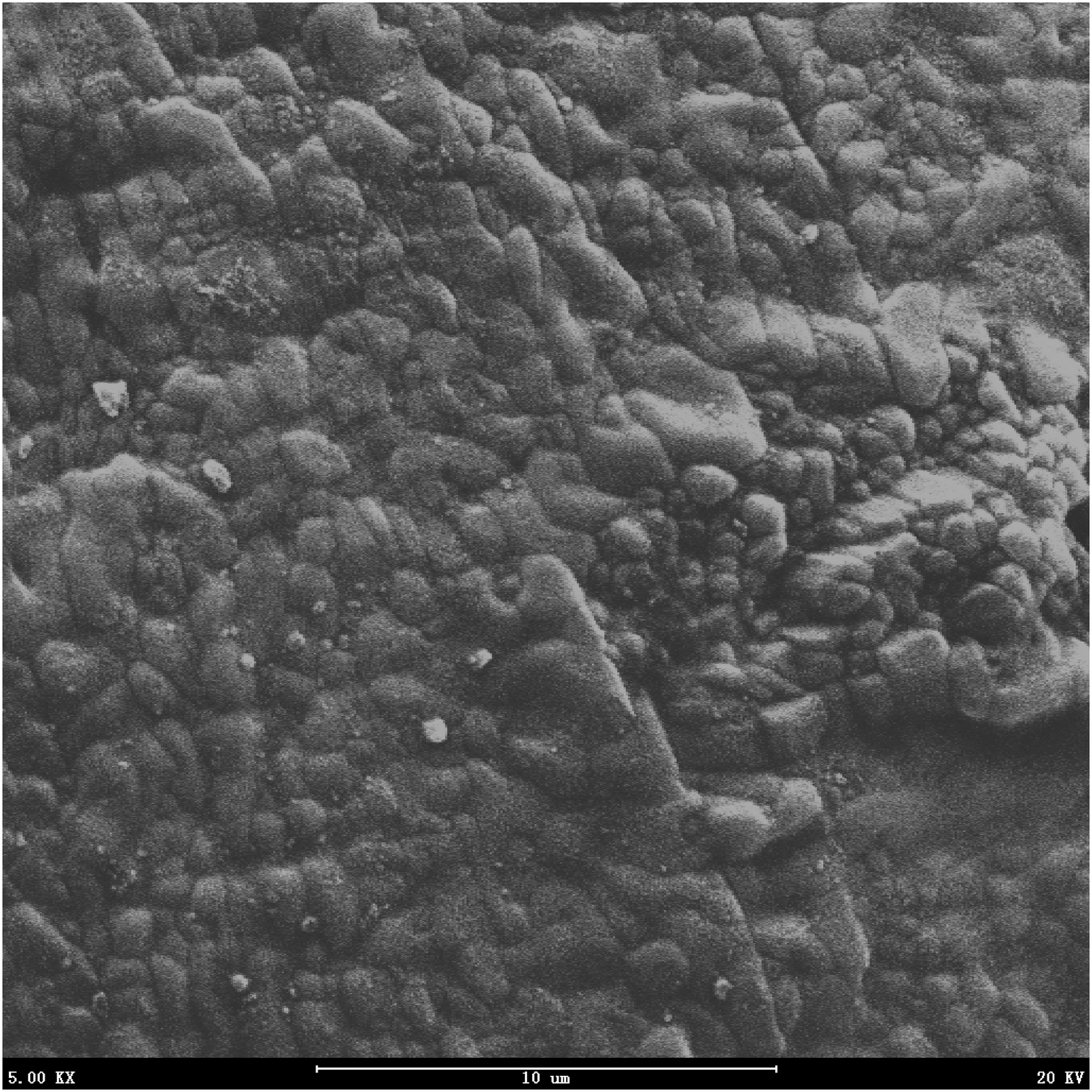}}
\caption{ SEM image of sintered \(CeO_2\) microspheres}
\label{fig_SEM}
\end{figure}

Scanning electron microscope (FE-SEM SUPRA 55) was used to investigate the morphology of microspheres. The two surface features deserving notification, sphericity and porosity are shown in the SEM images in Fig\ref{fig_SEM}. 

Judging from Fig\ref{fig_SEM1}\ref{fig_SEM2}, the sintered microspheres possess perfect sphere shape. And measurement yields that the sphericity of most of the microspheres is less than 1.05. A good sphericity of the microspheres is significant for in-core behavior in that spheres contributes to fuel isotropy, ease further blending in pellets fabrication process and help to transfer fission heat. Among all the geometry shapes, a sphere encloses the greatest volume for the same surface area. Therefore, the microspheres with a good sphericity transfer more heat through the same surface area from the greatest enclosed volume. 

Surface of a microsphere is magnified in Fig\ref{fig_SEM3}. The surface is not smooth and it can be attributed to the pores formed for further infiltration. In fact, the grain-formed surface implies sufficient interspace in the microsphere.

\begin{figure}
\centering
\includegraphics[width=0.4\textwidth]{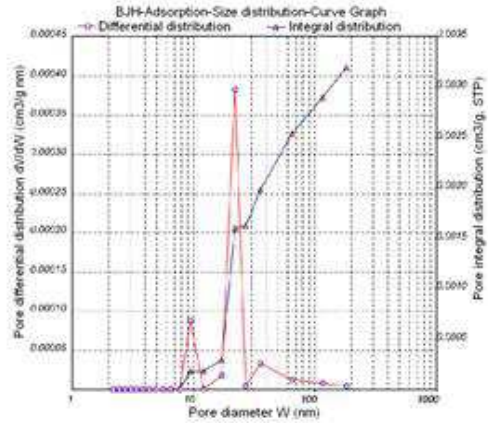}
\caption{Microsphere pore size distribution}
\label{fig_sr}
\end{figure}

The open pore size distribution of the sintered pellets was measured by a Surface Area and Porosimetry Analyzer (V-sorb 2800 ). And the result is given in Fig\ref{fig_sr}. The BET specific surface area is 0.070704. The pore size ranges from 8nm to 110nm and the distribution is peaked at about 20 nm. What is the best porosity for the further minor actinide infiltration is under research and will be published later. 

\section{Summary and conclusion}
A new method to fabricate ceric dioxide microspheres has been developed in this study as an analog preparation of plutonium ceramic nuclear fuel. A microfluidic device is applied to form gel microspheres containing cerium, the microspheres are then washed, dried, calcined, reduced and sintered during the fabrication process.

In the microfluidic device, the formed gel size can be controlled by adjusting material properties like viscosity of the continuous phase and procedure parameters including the flow rate of the phases. 

After being sintered, the fabricated microspheres turn out to be \(CeO_2\) particles possessing an expected small size with a narrow distribution and a good sphericity. The controllable sintered microspheres size ranges from \(10\nu m\) to \(1000\nu m\) and is distributed narrowly with a CV value less than 3\%. The pore size and its distribution are characterized and their optimization for further minor actinide infiltration is under investigation.

\section{Acknowledgement}
Authors are grateful to the \textbf{National Natural Science Foundation of China} for having funded this work through the grant No. \textbf{91226109} and No. \textbf{21076203}. Authors thank \textbf{National Cooperation Program of Anhui Province} for financial supporting in this research.

\bibliography{ExpR}
\end{document}